# Deep Learning Enables Large Depth-of-Field Images for Sub-Diffraction-Limit Scanning Superlens Microscopy

Hui SUN[1, #], Hao LUO[2,3, #], Feifei WANG[4,#], Qingjiu CHEN[1], Meng CHEN[1], Xiaoduo WANG[2,3], Haibo YU[2,3], Guanglie ZHANG[1,5,*], Lianqing LIU[2,3,*], Jianping WANG[5,6], Dapeng WU[5,6,*], Wen Jung LI[1,2,3,5,*]

**Scanning electron microscopy (SEM) is indispensable in diverse applications ranging from microelectronics to food processing because it provides large depth-of-field images with a resolution beyond the optical diffraction limit. However, the technology requires coating conductive films on insulator samples and a vacuum environment. We use deep learning to obtain the mapping relationship between optical super-resolution (OSR) images and SEM domain images, which enables the transformation of OSR images into SEM-like large depth-of-field images. Our custom-built scanning superlens microscopy (SSUM) system, which requires neither coating samples by conductive films nor a vacuum environment, is used to acquire the OSR images with features down to ~80 nm. The peak signal-to-noise ratio (PSNR) and structural similarity index measure values indicate that the deep learning method performs excellently in image-to-image translation, with a PSNR improvement of about 0.74 dB over the optical super-resolution images. The proposed method provides a high level of detail in the reconstructed results, indicating that it has broad applicability to chip-level defect detection, biological sample analysis, forensics, and various other fields.**

In many of the emerging developments in ultra-precision engineering, from the manufacturing of gigascale integration circuits to the development of micro- and nano-electromechanical systems, the scale of each studied device is between a few micrometers to a few nanometers. Conventional optical microscopes have diffraction-limited resolution, which is insufficient for humans to examine nanoscale objects with features smaller than ~200 nm. Overcoming the diffraction limit of current optical systems and obtaining efficient super-resolution methods can therefore contribute to fundamental research and to the advancement of nanoscale manufacturing. Accordingly, various super-resolution optical microscopes or microscopy techniques that can break the diffraction limit have been developed over the past two decades. The mainstream microscopy systems or techniques can be broadly classified into four categories based on the imaging principle used: fluorescence-based super-resolution microscopy[1], surface plasmon polaritons[2] (SPPs), structured light illumination–based super-resolution microscopy[3], and microsphere lens–based super-resolution microscopy[4].

Stimulated emission depletion microscopy (STED), structured illumination microscopy (SIM), photo-activated localization microscopy (PALM), and super-resolution optical fluctuation imaging (SOFI) are the four basic fluorescence-based approaches to achieve super-resolution imaging. Hell introduced STED[5] in 1994 and the approach suppresses fluorescence emission by modulating the excitation beam and reducing the effective fluorescence point spread function, which is based on optical nonlinear effects. SIM[6] is an ultra-high resolution method that was presented by Gustafsson in 2000 and is based on the principle of structured light illumination. This technology achieves high-resolution imaging by modifying the light beam using moiré fringes and acquiring high spatial–spectral information through image reconstruction. In 2006,

1. Department of Mechanical Engineering, City University of Hong Kong, Hong Kong SAR, China.
2. State Key Laboratory of Robotics, Shenyang Institute of Automation, Chinese Academy of Sciences, Shenyang 110016, China.
3. Institutes for Robotics and Intelligent Manufacturing, Chinese Academy of Sciences, Shenyang 110169, China.
4. Department of Electrical and Electronics Engineering, The University of Hong Kong, Hong Kong SAR, China.
5. City University of Hong Kong Shenzhen Research Institute (CityUSRI), Shenzhen 518000, China.
6. Department of Computer Science, City University of Hong Kong, Hong Kong SAR, China.
\# These authors have contributed equally to this work.
\* CORRESPONDING AUTHORS: Guanglie Zhang; Lianqing Liu; Dapeng Wu; Wen Jung Li (e-mail: gl.zhang@cityu.edu.hk; lqliu@sia.cn; dapengwu@cityu.edu.hk; wenjli@cityu.edu.hk)

three stochastic reconstruction techniques that combine single-molecule localization with fluorescent molecular switching effects were proposed almost simultaneously: stochastic optical reconstruction microscopy (STORM) by Zhuang et al.[7], PALM by Betzig et al.[8], and fluorescence PALM by Hess[9]. Subsequently, SOFI[10] was proposed by Dertinger et al. in 2009. To accomplish high-resolution imaging, this approach takes advantage of the variations in the intensity of continuous luminescence of fluorescent molecules over time and executes a higher-order time-dependent operation by the cumulative detection of this intensity. Of these fluorescence-based super-resolution methods, STED, STORM/PALM, and SOFI are more reliant on fluorescent molecular properties than SIM. The limits of fluorescence super-resolution methods are therefore attributable to the properties of fluorescent molecules. To some degree, their broad usage has been hampered by the high cost of the equipment and the difficulty of tagging many materials successfully with fluorescent molecules.

In 2011, Wang et al.[11] introduced the microsphere-based super-resolution imaging method by proposing a microsphere lens imaging technique in which transparent dielectric silicon dioxide microspheres with diameters of 2–9 μm are placed on the sample surface, and images with a resolution of 50 nm can be obtained under a white light source using a conventional optical microscope, enabling super-resolution optical imaging. This straightforward and efficient technology opened new possibilities for real-time, in-situ super-resolution optical imaging. In 2015, another group (Wang et al.[12]) attached dielectric microspheres to atomic force microscopy (AFM) probes to provide 3-axes nanometer scale movement control of microsphere lenses while scanning across sample surfaces. The super-resolution images observed in the scanned area of the microsphere are tracked and recorded while the microsphere is in motion, and the captured images are stitched together and processed to achieve large field-of-view super-resolution imaging. However, the super-resolution images acquired by scanning-microsphere microscopy do not have high-contrast and large depth-of-field when compared with the scanning electron microscopy (SEM) images.

On the other hand, the use of SEM can overcome optical diffraction limitations and enable micro- or nanostructure imaging, but increasing the spatial resolution of pictures using SEM has some drawbacks in the increased size and weight of the imaging equipment and optical components, the complexity involved in manufacturing and processing optical materials, and the high cost of usage. As the operation of SEM requires the sample to be placed in a vacuum environment, the sample must be dry, oil-free, and conductive of electricity, which cannot be achieved for applications involving the observation of living biological material. It is often impractical to depend only on hardware advancements to produce high-quality high-resolution pictures, as they are affected by variables such as process level and various sample restrictions and are frequently prohibitively expensive. To solve the limitations related to physical conditions, therefore, deep learning algorithms for super-resolution image reconstruction can realize the transformation of optical super-resolution images into SEM images at the software level.

Unlike the optical super-resolution technique, the image super-resolution technique is a software algorithm to improve image resolution, which uses deep learning methods to reconstruct a high-resolution (HR) image from the input single or multiple low-resolution (LR) images such that the super-resolution image contains high-frequency details that do not exist in the LR image. The super-resolution reconstruction technique can scale up the image and increase its size, and has features such as deblurring, denoising, and restoring high-frequency details of the image. In various image transformation tasks[13–15], deep learning has resulted in substantial advances by attempting to recover HR images from their LR counterparts. Among the super-resolution deep learning networks that have emerged in recent years are Super-Resolution Convolutional Neural Network (SRCNN)[16], Efficient Sub-Pixel Convolutional Neural Network (ESPCN)[17], Very Deep Convolutional Networks for Super-Resolution (VSDR)[18], Enhanced Deep Residual Networks for Super-Resolution (EDSR)[19], Deeply-Recursive Convolutional Network (DRCN)[20], Laplacian Pyramid Super-Resolution Network (LapSRN)[21], Deep Recursive Residual Network (DRRN)[22], SRDenseNet[23], Super-Resolution Generative Adversarial Network (SRGAN)[24], Persistent Memory Network (MemNet)[25], Multi-Channel Dense Connections (MCRAN)[26] and Residual Dense Network (RDN)[27]. The structure of these deep learning networks mainly consists of feature extraction, detail prediction, and reconstruction output. The first super-resolution deep learning network was SRCNN, which uses a three-layer network to simulate the conventional super-resolution method[28]. Following that, ESPCN was proposed. It uses pixel rearrangement for extracting features from LR images by directly performing convolution operations on the LR images and then expanding the features in LR space into HR space using sub-pixel convolution layers [17]. The rebuilt image is created by a pixel-wise arrangement of the characteristics acquired after convolution. The reconstruction effect is far superior to the interpolation technique because it does not solely utilize neighboring pixels to compute the pixel values to be interpolated. At the same time, by changing the number of feature channels flexibly, this technique can rapidly achieve various magnifications, offering a novel strategy for obtaining different magnifications of images at super-resolution. Kim et al.[18] were the first to integrate a residual structure into the super-resolution technique, resulting in the VDSR model with a 20- layer depth. VDSR outperforms SRCNN by using several layers of tiny convolutional kernels for deep convolution, which lowers the number of parameters while increasing the deep network's perceptual field and improving its feature learning. Furthermore, given that both HR and LR images include a significant quantity of identical low-frequency information, utilizing the residual structure may prevent the need for learning similar LR information several times, which speeds up network convergence and saves computation time. However, with just a one-skip connection introduced in this network, the gradient vanishing issue is not well addressed. The EDSR network model was proposed by the SNU CV Lab team in 2017[19]. It uses a deep residual network to alter the network structure and removes the batch normalization layer, which not only saves memory but also speeds up the calculation. Furthermore, EDSR uses a data enhancement strategy (called a geometric self-ensemble), in which each LR image is geometrically transformed several times to obtain the corresponding HR image, with the results then averaged to obtain the final reconstructed image; this improves the network's stability. In 2018, Zhang et al.[29] developed the RCAN model, bringing the channel attention mechanism in signal processing to the super-resolution field. The channel attention mechanism improves both the network's ability to emphasize relevant channels and its discriminative learning ability.

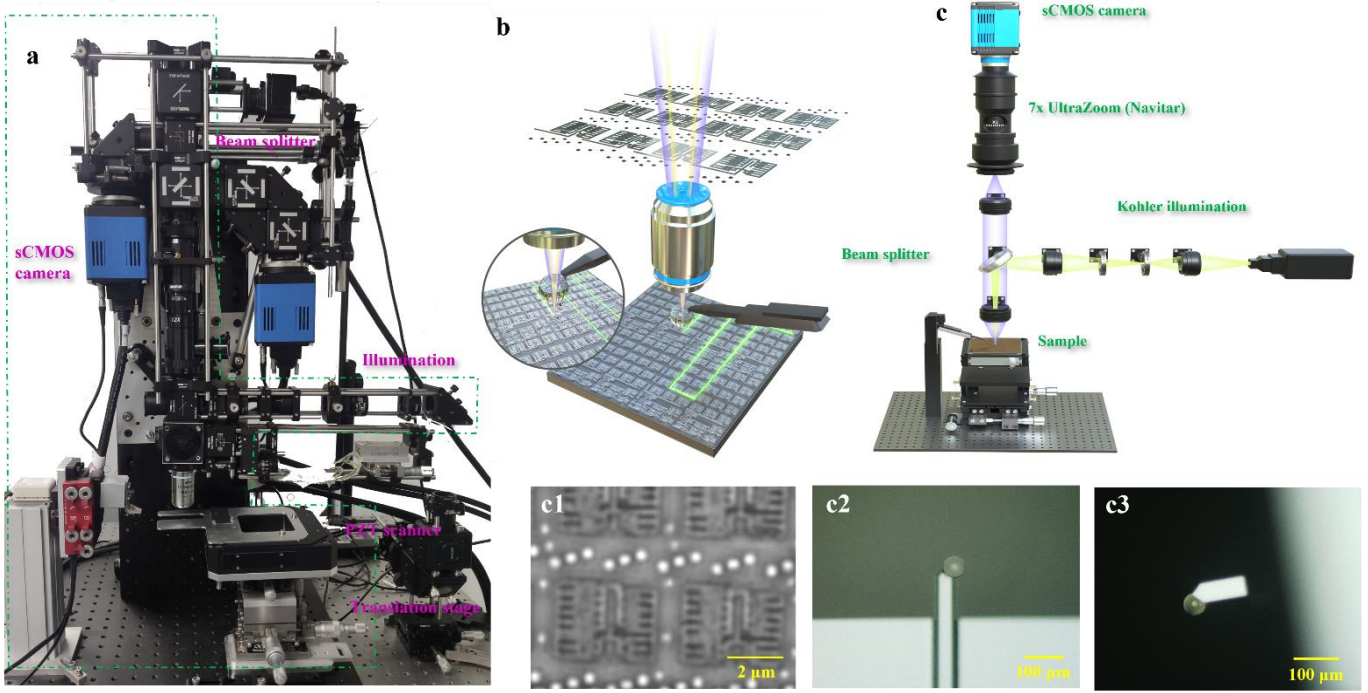

**Fig. 1 | Scanning superlens microscopy (SSUM). a** Photograph of the SSUM setup. **b** Schematic of a microsphere-based SSUM that integrates a microsphere superlens into an AFM scanning system by connecting the microsphere to an AFM cantilever. The objective collects virtual images containing sub-diffraction-limited object information while also focusing and collecting the laser beam utilized in the cantilever deflection detection system. **b1** An original virtual image observed using the microsphere superlens. The inset shows an SSUM image. **b2** and **b3** Front-side and rear-side images, respectively, of the AFM cantilever with an attached microsphere superlens. Scale bars: 2 μm (**b1**); 100 μm (**b2, b3**). **c** Schematic diagram of the optical path of the SSUM system.

While the above mentioned deep learning models can be used to improve the resolution of scientific microscopy images, these single-image super-resolution (SISR) methods do not enable a more accurate and quantitative inference of the nanostructure of the object. As a result, further image resolution enhancement based on optical super-resolution is required to achieve super-resolution images with an SEM resolution effect. In this paper, we design a scanning superlens micropcopy (SSUM) system (as shown in **Fig. 1a**) to overcome the optical diffraction resolution limit. We combine this microsphere-assisted super-resolution imaging system with the Cycle-consistent Generative Adversarial Network (CycleGAN) deep learning method to learn the mapping relationship between the optical images and SEM images, which allowsfor more accurate super-resolution imaging. As depicted in the assembly schematic in **Fig. 1b** and the system imaging schematic in **Fig. 1c**, we design a probe-lens assembly optical super-resolution imaging system using barium titanate glass BTG microspheres, a three-dimensional (3D) translation stage, and a scanning platform that can freely investigate the sample surface and accomplish optical super-resolution imaging with a large field-of-view. To further improve the resolution of optical images to SEM images, a deep learning algorithm is used to achieve image super-resolution reconstruction. The reconstruction process of the CycleGAN with multi-residual block-based image super-resolution technique is shown in **Fig. 2**. We elaborate on the design of the network structure **in the *Method* section** and the selection of the loss function for deep learning **in the Supplementary Fig. S4**. Finally, we analyze and compare the network output images and the reference images (SEM images) in the spatial and frequency domains, and quantify our results using the peak signal-to-noise ratio (PSNR) and the structural similarity index measure (SSIM) **(see Supplementary Fig. S1)**. The CycleGAN method shows a significant improvement over the conventional method in image clarity and detail enhancement, and the obtained optical super-resolution image reconstruction results are more detailed and natural in terms of visual effects.

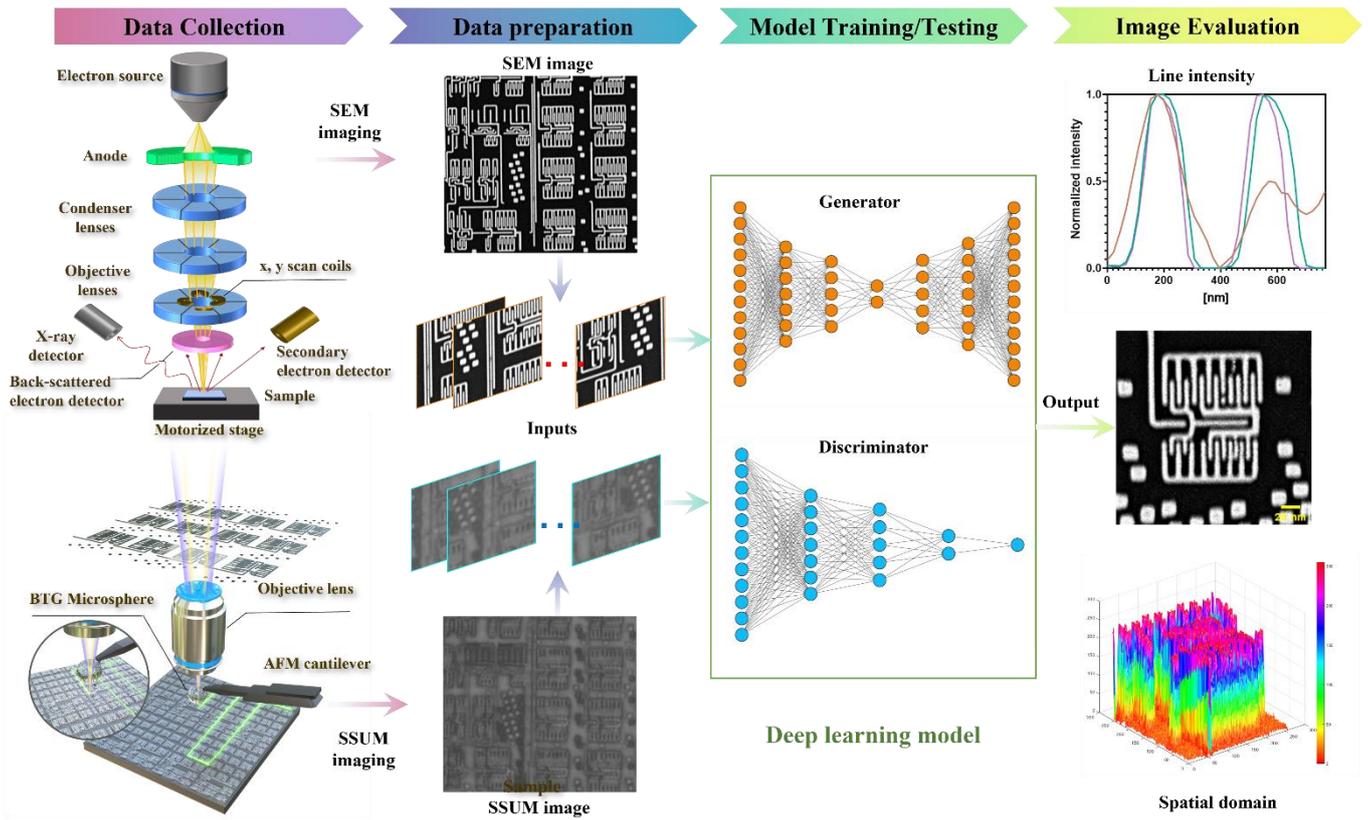

**Fig. 2 | Schematic diagram of image super-resolution reconstruction for the proposed optical SSUM system using the reference-based super-resolution algorithm.**

## Results

### Comparison in spatial domain

The imaging quality of the generated super-resolution images improves as the number of training iterations increases using the deep learning method. At 500 epochs, both the detail and contrast of the images are also significantly enhanced. CycleGAN with multi-residual block-based image super-resolution reconstruction can produce high-resolution images, as shown in **Fig. 3**. The neural network model can quickly train images with high resolution because it is trained on the basis of residual blocks. The grayscale distribution of the image corresponds to the intensity magnitude of each pixel, and can be used to represent global similarity between images to some degree. **Fig. 3f–h** show the global grayscale distribution of optical super-resolution images, super-resolution reconstructed images based on CycleGAN with multi-residual blocks, and SEM images, respectively. The results show that the super-resolution reconstructed image based on deep learning is closer to the real SEM image in terms of the global grayscale distribution, which is not only a visual enhancement, but more importantly, the results generated using the super-resolution reconstruction method proposed in this paper are also fidelity and credible.

### Comparison in frequency domain

As the spatial-domain signal of an image can only reflect changes in the amplitude of the signal, it is difficult to analyze the composition of the information frequency and the magnitude of the frequency components in detail, except for the simple harmonic of a single frequency component. The signal spectrum, however, provides much richer information. A signal's frequency domain distribution is determined by the Fourier transform of an image. For discrete signals like digital images, the frequency magnitude indicates how drastically the signal changes or how fast the signal varies. The higher the frequency, the more dramatic the shift, and the lower the frequency, the flatter the signal. As a result, the high-frequency signal commonly corresponds to the image's edge signal and noise signal, and the low-frequency signal comprises the image contour and background signal in situations when the image changes rapidly. **Fig. 3i–k** show the results of the Fourier transform and frequency spectrum centering of optical super-resolution images, deep learning-based image super-resolution reconstructed images, and SEM images, respectively.

The low-frequency component of the image has high energy and is distributed in the center of the image, as seen in **Fig. 3i–k**, whereas the high-frequency characteristics of the image are dispersed on its two sides. Compared with the super-resolution images generated using the multi-residual-block CycleGAN and with the SEM images, the optical super-resolution images exhibit fewer high-frequency features, resulting in a blurred image. In terms of high-frequency distribution, the super-resolution image produced with the deep learning method is highly similar to the SEM image; hence, the generated effect is very close to the real image.

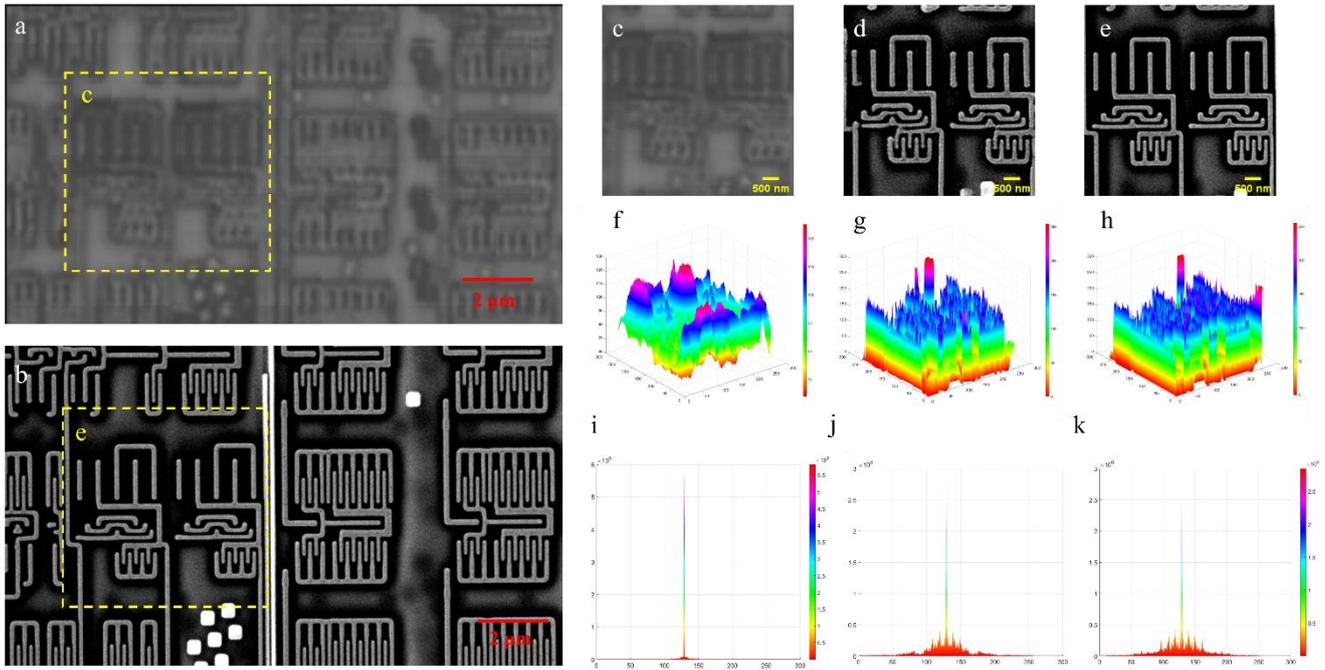

**Fig. 3 | Comparison of image super-resolution reconstruction results.** From the SSUM image **a** and the SEM reference image **b**, a region **c** is selected for testing the super-resolution model. The optical super-resolution image taken by SSUM is used as the input to the model, and the reconstructed super-resolution image **d** is generated after the deep learning-based super-resolution model, with **f** as the local reference image. In the spatial domain, **f–h** are the grayscale distribution results at different locations for optical super-resolution images, CycleGAN with multi-residual block-based image super-resolution reconstructed images, and SEM images, respectively. In the frequency domain, **i–k** show the results of Fourier transform and frequency spectrum centering of optical super-resolution images, CycleGAN with multi-residual block-based image super-resolution reconstructed images, and SEM images, respectively.

**Line profile analysis**

A wide range of areas on the silicon wafer sample were inspected using the microsphere-assisted optical super-resolution imaging method, as shown in **Fig. 4a** and **Fig. 4h**, corresponding to the same areas observed using a 100× objective, as shown in **Fig. 4b** and **Fig. 4i**. Two areas were then randomly selected and compared between images observed with a 100×/0.90 objective, optical super-resolution imaging with microspheres, deep learning-based super-resolution imaging, and SEM. We carried out a line profile analysis (**Fig. 4g** and **Fig. 4n**) of the horizontal line through the "gap" between its neighboring lines. The results for these four images are shown in **Fig. 4c–f**, and those for normalized light intensities are in **Fig. 4 j–m**. It is obvious from the results that the imaging results are least clear when using an optical microscope with a 100× objective, and the details of the object under the optical diffraction limit cannot be seen. SSUM imaging technique can break the optical diffraction limit, but cannot further improve the resolution of the image. For the super-resolution reconstructions based on deep learning, the intensity profiles of the images are close to those observed from SEM images. The smallest structural unit in this silicon wafer we used is 73 nm, which can be well observed. These results show that the resolution using microspheres exceeds that of conventional optical lenses and that combining super-resolution image reconstruction using deep learning with optical super-resolution images can achieve results that are very close to those of SEM imaging. Our proposed method can make the generated image features clearer, with less noise interference and better visual effect.

**Quantification of super-resolution artifacts using NanoJ-SQUIRREL**

The resolution is typically estimated by measuring the minimum resolvable distance between two adjacent structures in the image. The Fourier ring correlation (FRC)[30] method of measuring effective image resolution is a straightforward and objective approach. FRC measures the degree of correlation of two images at different spatial frequencies and is a prominent approach for assessing image resolution in super-resolution and electron microscopy. To determine the resolution threshold (the spatial frequency) at which both reconstructions are consistent, FRC compares the similarity of two independent reconstructions of the same object in frequency space. We used block FRC resolution mapping to obtain local resolution measurements in the NanoJ-SQUIRREL package. The images were then spatially segmented into equal-sized blocks, as shown in **Fig. 5d**, **Fig. 5h**, **Fig. 5l**, and **Fig. 5p**, and an FRC analysis was performed on each block, with **Fig. 5d** and **Fig. 5l** subdividing the image into equal blocks of size 10 and **Fig. 5h** and **Fig. 5p** subdividing the image into equal blocks of size 5. **Fig. 5** shows the resolution and quality of the images for different types and different areas, indicating that the blocks with lower FRC values have higher resolution. There may be some non- square blocks on the map; this happens when the correlation between the two images at this point is insufficient to compute the FRC value. The FRC values in these blocks are tessellated from neighboring blocks to adjust for this. In addition, other quantitative image performance indicators, i.e., PSNR and SSIM, were selected to assess the quality of the generated super-resolution images. Twelve images were randomly selected for testing, with the results shown in the **Supplementary Fig. S1** and **Table S1**.

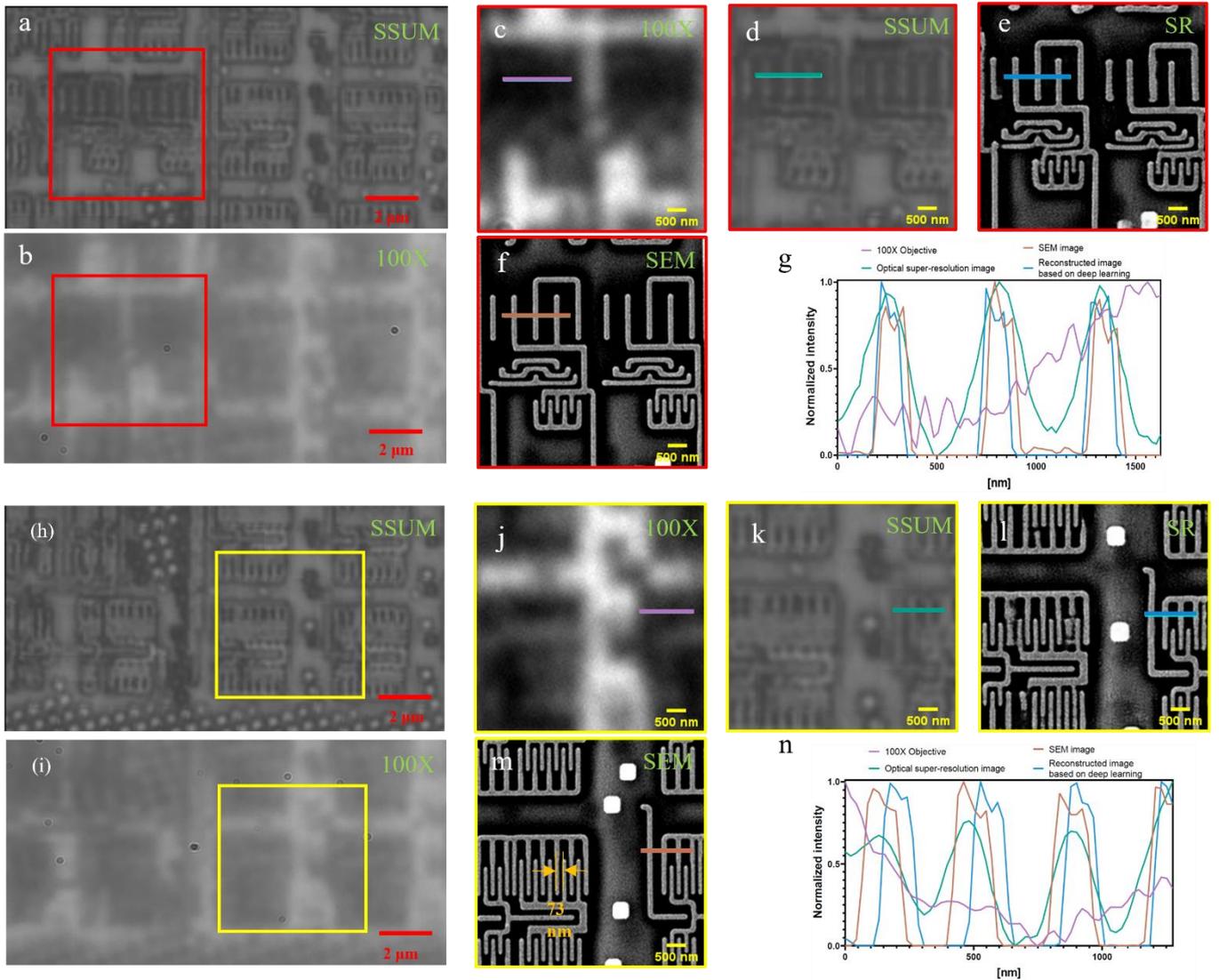

**Fig. 4 | Detected smallest features. a** and **h:** Optical super-resolution images of a silicon wafer; **b** and **i**: The same areas observed with the 100× objective corresponding to **a** and **h**; Location A in **c–f** and Location B in **j–m** are zoomed-in images of the red rectangular area and the yellow rectangular area, respectively; **c** and **j**: 100×-in-air; **d** and **k**: Optical super-resolution with microsphere; **e** and **l**: Super-resolution reconstruction based on deep learning; **f** and **m**: SEM images; **g** and **n**: Line profiles of the blue dashed lines in **c–f** and **j–m** with normalized intensity.

**Comparison of different super-resolution methods**
Super-resolution images produced by three image super-resolution algorithms, Super-Resolution Generative Adversarial Network (SRGAN)[24], Enhanced Super-Resolution Generative Adversarial Networks (ESRGAN)[31], and Pix2Pix[32], are compared in **Fig. 6**. The low-resolution input images in the original algorithm were generated by down-sampling the ground truth images, and the high-resolution image and low-resolution image were replaced by an SEM image and optical super-resolution image, respectively, as the training model input. Because SRGAN and ESRGAN are both SISR algorithms, learning the image-to-image mapping relationship between low-resolution and high-resolution images is difficult, resulting in unsatisfactory image generation. Although Pix2Pix is a reference-based super-resolution algorithm, it overemphasizes the one-to-one mapping relationship between image pairs, and because the SEM images and optical super-resolution images are matched manually during the dataset production, some image matching errors are inevitably generated and gradually accumulate over a series of image cropping and other operations, thus limiting the effect of the algorithm. Therefore, we adopted the CycleGAN algorithm, which has relatively lenient image matching requirements, to realize the conversion of different types of images and thus achieve the generation of SEM super-resolution images. From the results in the **Fig. 6**, the super-resolution images obtained using the CycleGAN algorithm are closest to the real images.

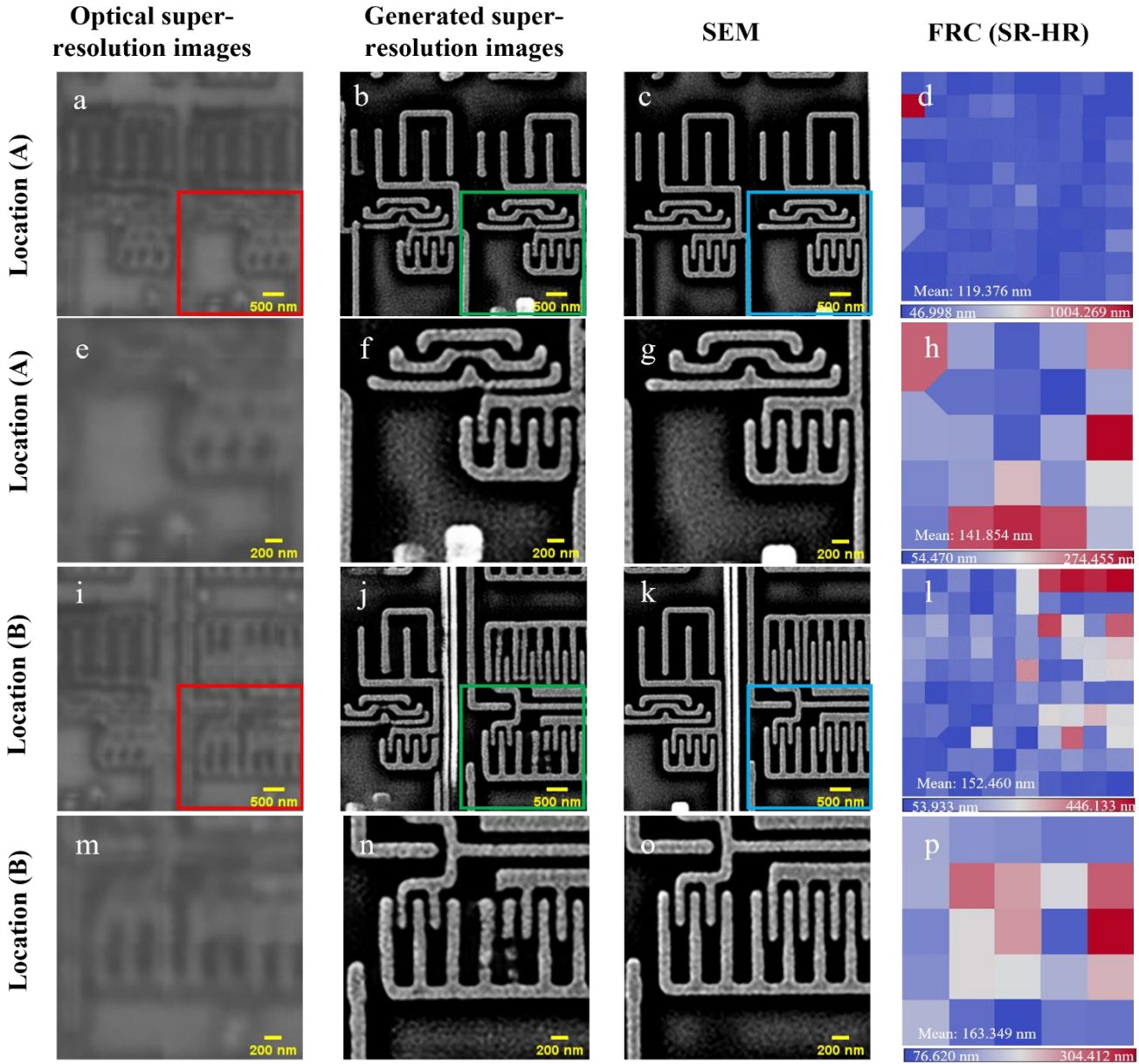

**Fig. 5 | Image super-resolution imaging results based on CycleGAN with multi-residual blocks and resolution mapping with NanoJ-SQUIRREL. a–c** and **i–k**: Optical super-resolution imaging; **b** and **j**: Super-resolution imaging results based on CycleGAN with multi-residual blocks; **c** and **k**: SEM imaging (reference); **e–g**: Zoomed-in images labeled in **a–c**; **m–o**: Zoomed-in images labeled in **i–k**; **d**, **h**, **l** and **p**: SQUIRREL resolution map of the super-resolution image between the generated super-resolution images and SEM images. Scale bar: 500 nm for **a–c** and **i–k**; 200 nm for **e–g** and **m–o**.

**Super-resolution reconstruction of images with a larger field of view**

To obtain super-resolution images with a larger field of view, we performed deep learning-based super-resolution image reconstruction on optical images of silicon wafers with an actual physical size of 18.5 μm × 18.5 μm. The results shown in **Fig. 6i–k** indicate that using the image super-resolution method in this paper, the SEM super-resolution images can be transformed very well, not only for optical images with a small field of view but also for those with a large field of view.

**Discussion**

The optical super-resolution system based on deep learning presented in this article has the following key functions. First, for silicon wafer image capture, this system uses an optical microscope in combination with a microsphere-lens super-resolution imaging technique, which overcomes the optical diffraction limit and allows for a more detailed examination of the structure of nanoscale silicon wafers. Although this optical microscopy super-resolution technology overcomes the optical diffraction limit, it does not fulfill the requirements for high-resolution images and does not reach the imaging effect of SEM images. Second, therefore, a generative adversarial network based on optical microscopy super-resolution images is proposed to transform low-resolution optical microscopy images to super-resolution SEM images, thus overcoming the limitations of the optical device manufacturing process and manufacturing cost. A multi-residual-block CycleGAN super-resolution prediction and reconstruction method based on registered microscope image pairs is proposed, allowing researchers to observe nanoscale silicon wafer structures with lower resolution, faster imaging speed, and resolution beyond the diffraction limit, and to achieve the super-resolution reconstruction of images across optical and SEM platforms.

The produced super-resolution images were compared and analyzed in both the spatial and frequency domains. The grayscale distribution of the generated super-resolution images is very similar to that of the real SEM images, and the frequency distribution of the generated super-resolution images has more high-frequency components, indicating that the images contain more detailed information. On the image evaluation metrics of pixel-wise difference, brightness, contrast, and structure of the images, the PSNR and SSIM values of the images generated using the optical microsphere-lens super-resolution method based on deep learning are higher than those of the images generated using only optical microsphere-lens super-resolution, indicating that the quality of the generated super-resolution images is closer to that of the real SEM images.

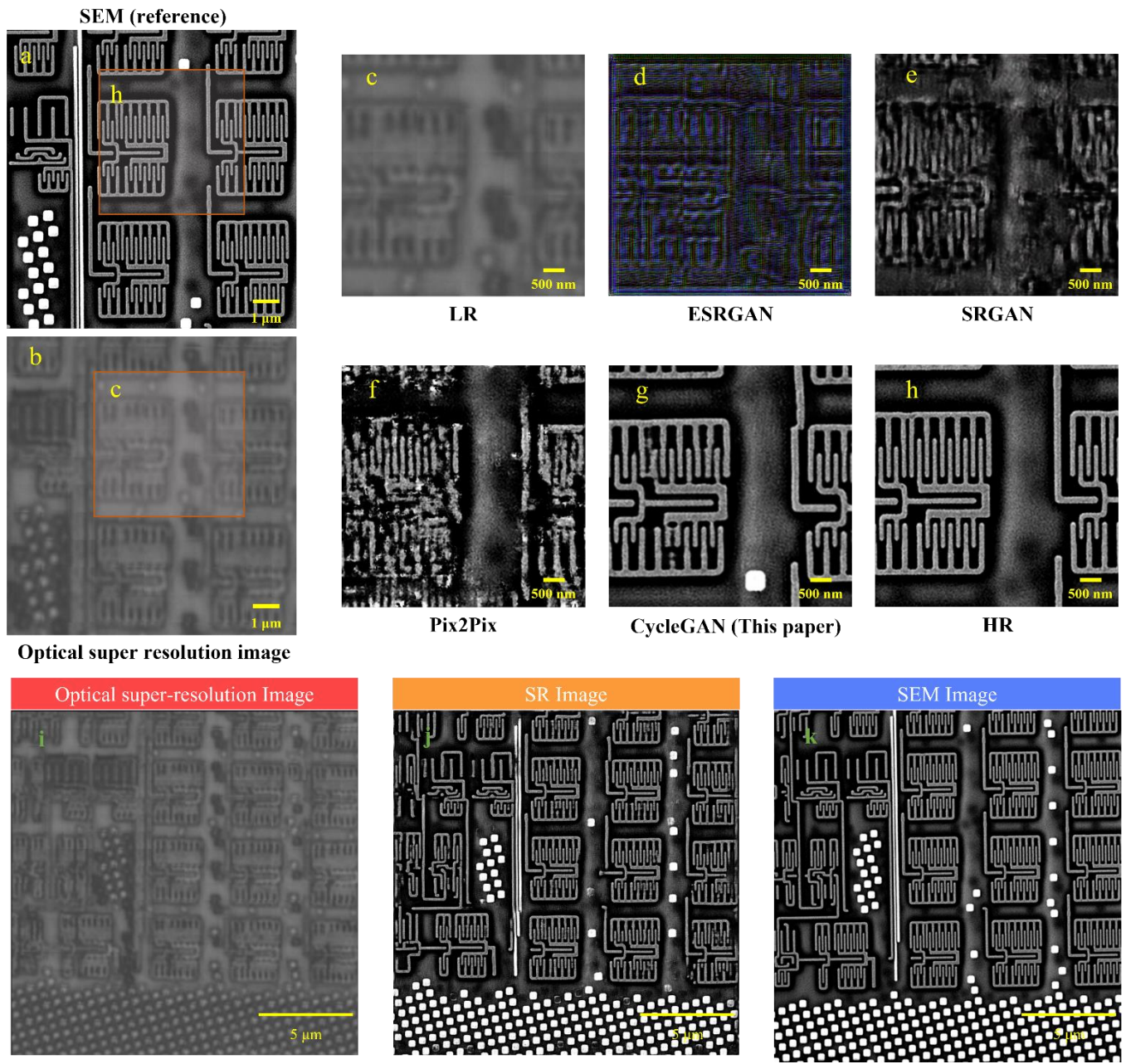

**Fig. 6 | Comparison of super-resolution methods and the reconstruction super-resolution image with a large field of view.** Area **h** and **c** were chosen from the optical super-resolution image in **b** and the SEM image in **a**, respectively, and several super-resolution image reconstruction methods applied to the area in **c**. The image super-resolution reconstruction techniques based on ESRGAN, SRGAN, Pix2Pix, and CycleGAN are shown in **d**–**g**, respectively. Scale bar: 500 nm. **j** is a super-resolution reconstruction image with a large field of view based on CycleGAN. The reference image is represented by **k**, and the image to be reconstructed by **i**. Scale bar: 5 μm.

To demonstrate the multi-scale super-resolution effect of the super-resolution model and the robustness of the model, we tested SSUM images with a large field of view, with the results shown in **Fig. 6j**. Even with the larger field of view, the super-resolution reconstruction of the images is effective. To verify the robustness of the super-resolution model, some more complex SSUM images were also selected for tests. The **Supplementary Fig. S2** shows that the model is still able to learn the complex point and line features well to achieve a super-resolution reconstruction of the image.

**Methods**

**Principle and design of optical super-resolution system**

Based on our previous work[12,33], we propose a non-invasive, high-throughput, environmentally compatible optical SSUM system, the composition and performance characteristics of which are shown in the **Supplementary Fig. S3**. **Figure 1** shows the schematic construction of the SSUM, which can be used for large-area, super-resolution imaging and data acquisition. The optical super-resolution imaging system consists of an AFM, a commercially available cantilever (TESP probe, Bruker), a 57-μm-diameter BTG microsphere lens (Cospheric) and a 100× objective lens (Nikon LU Plan EPI ELWD). Microspheres are attached to the AFM probe cantilever using a UV-curable adhesive (NOA63, Edmund Optics). Different illumination conditions are achieved by adjusting two stops in the Köhler illumination system (Thorlabs). A high-speed scientific complementary metal oxide semiconductor camera (PCO.Edge 5.5) is used to record the images. Illumination is provided by an intensity-controlled light source (C-HGFI, Nikon, Japan), with the peak illumination wavelength of the system set to ~550 nm for white light imaging by the optical elements.

A drop of UV-cured adhesive is positioned close to the microsphere region to act as the sticky material. The selected microspheres are touched by the adhesive-coated tip, which is then subjected to UV light for 90 seconds until the adhesive is fully cured. The microsphere superlens is mounted to the AFM cantilever and kept in place during scanning to keep the distance between the microsphere and the objective as consistent as possible. The specimen is placed on top of the stage after the standard AFM adjustments are completed. The scanning platform and 3D translation stage are used to carry out horizontal sample scanning and vertical focus adjustment. The IC chip scanning in the horizontal direction and the feedback adjustment in the longitudinal direction are realized by a 3D piezoelectric ceramic (PZT) scanner (P-733.3CL, Physik Instrumente, Germany), in the process of which the AFM feedback is acquired to monitor whether the microspheres are touching the sample surface and to adjust the distance between the microspheres and the sample. When the distance or force reaches a certain value, the optical microscope is driven by a translation stage with nanoscale resolution (IMS100V, Newport) to capture a virtual image generated by the microsphere. The interval of the IC chip scan and the interval of the signal used to trigger the camera image recording are both adjusted according to the region of the field of view of the microsphere superlens during horizontal scanning. Before scanning, the camera's captured area can be modified to satisfy the overlap requirement for image stitching, and there is no considerable aberration in the field of view of the microsphere superlens, which effectively reduces data processing time and allows for fast image processing before or after scanning.

**SEM image acquisition**

The HR images are captured using a Quanta$^{TM}$ 450 FEG-SEM (field-emission gun scanning electron microscope; SEI) with a beam voltage of 30 KeV and a spot size of 4.0.

**Image pre-processing**

To enable the model to better learn the mapping relationship between optical super-resolution images and SEM images, image processing of the captured optical images and SEM images is necessary. To gather the optical super-resolution(low-resolution) and SEM (high-resolution) image pairs of the same samples for network training, it is first necessary to stitch together each frame image taken by the optical super-resolution device to obtain an optical super-resolution image of the wafer. After that, the optical super-resolution image is registered with the SEM image, matching the fields of view of the LR and HR images so that each training pair shows roughly the same region of the sample, and cropping them to 512×512 pixels. If the effective pixel sizes of the LR and HR images are different (e.g., the images are captured with different devices, the SSUM system and SEM), we rescale them to the same physical size. The cropped image pairs are then sorted into two groups for use as training and test datasets, with the training dataset being fed into the neural network for model training. In terms of network structure design and loss function, a network model is built by combining convolutional neural networks and multiple residual blocks, and the loss function is designed based on prior knowledge; model training is then performed to determine the optimizer and learning parameters, and the network parameters are updated using a backpropagation algorithm to improve the model's learning ability by minimizing the loss function. The network model is evaluated according to the performance of the trained model on the validation set and corresponding adjustments are made. Finally, the trained generative model is tested with the test dataset images to evaluate the tested image quality **(see the Supplementary Fig. S1 and Table S1)**.

**Design of CycleGAN with multi-residual block network architecture**

CycleGAN is a method for training deep convolutional neural networks using unpaired datasets for image-to-image translation tasks, with the network structure and loss function shown in **Supplementary Fig. S4**. The basic goal of this translation is to use a network to learn a mapping between input and output images. During the training phase, cycle consistency is needed: that is, one image identical to the original should be produced with the lowest L1 loss value following repeated application of two different generators. **Supplementary Fig. S5** shows the variation in error for the generator and discriminator in the image super-resolution model, after numerous iterations.

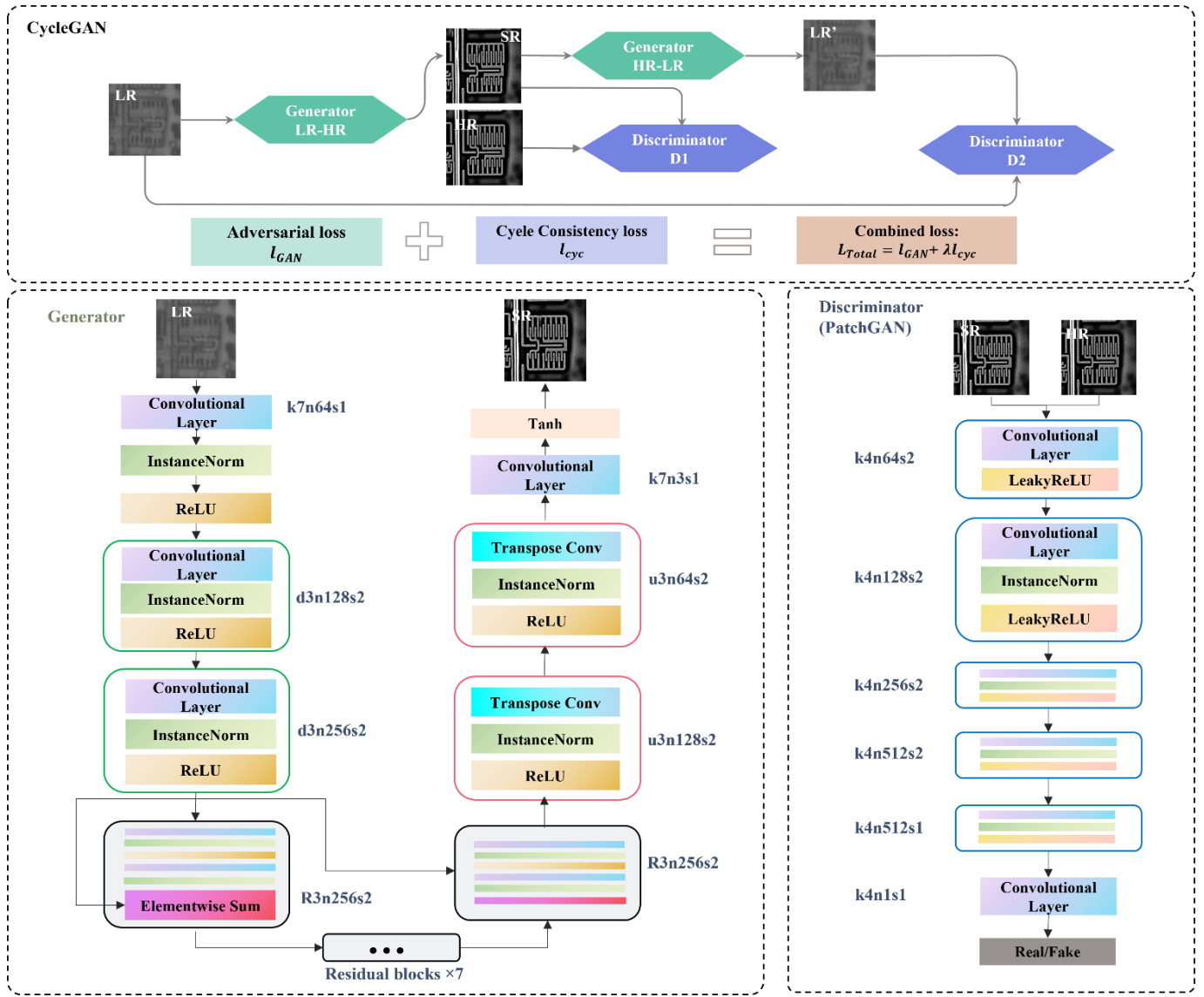

**Fig.7 | Structure of image super-resolution network based on CycleGAN with multi-residual blocks.**

**Network architecture**
The network structure of the CycleGAN with multi-residual blocks is shown in **Fig. 7**. Its main principle is to train the generator and discriminator models to convert images from low to high resolution with cyclic consistency, such that the generated super-resolution image should subsequently be back-converted to the LR image. We allow a generator to learn to transform features of synthetic images into features similar to those in a real image, while a discriminator learns to distinguish real from synthetic images. For the discriminator networks, we use 70×70 PatchGANs, which aim to classify whether each patch in the image is real or fake. A patch-level discriminator architecture has fewer parameters than a full image discriminator and can be applied to arbitrarily sized images in a fully convolutional fashion. For each training batch, the generator and the discriminator compete against each other so that the generator learns to produce features sufficiently similar to the real image to fool the discriminator, bringing the generated super-resolution images closer to the real high-resolution images.

**Generator architecture**
The generator contains three main modules: the downsampling module, the residual module, and the upsampling module. During training, the optical super-resolution image is downsampled to an LR image. The generator architecture then attempts to upsample the image from LR to super-resolution, and shallow features are extracted using convolution. The residual blocks are then used to extract the deeper features. The generator architecture then attempts to upsample the image from low resolution to super-resolution using the upsampling module. After this, the image is passed into the discriminator, which tries to distinguish between the SEM image and generated super-resolution image and generate the adversarial loss for backpropagation into the generator architecture.

In the network structure diagram in **Fig. 7**, c, d, R, and u represent the convolution kernel size of the convolution layer, downsampling layer, residual block, and upsampling layer, respectively. To extract residual characteristics from both LR and HR images, a deep learning network structure is built with nine residual blocks for 256×256 resolution training images. The generator architecture contains residual networks instead of deep convolutional networks because residual networks are easier to train and can thus be substantially deeper to generate better results. The benefit arises from the residual network using a type of connections called skip connections. There are nine residual blocks, generated by ResNet. Within each residual block, two convolutional layers are used with small 3×3 kernels and 64 feature maps followed by batch normalization layers and LeakyReLU as the activation function. The resolution of the input image is increased using two trained sub-pixel convolution layers. The above mentioned network structure can adaptively learn the parameters of the rectifier and improves the accuracy at negligible extra computational cost.

**Discriminator architecture**
The task of the discriminator is to discriminate between real HR images and generated super-resolution images. The discriminator architecture used in this paper is a 70×70 PatchGAN architecture with LeakyReLU as the activation function. The structure also uses instance normalization. The network contains five convolutional layers. With the deepening of the network layers, the number of features increases and the feature size decreases. The first four layers have 4×4 filter kernels, increasing by a factor of 2 from 64 to 512 kernels with stride 2, which is the convolutional-InstanceNorm-LeakyReLU structure. The last layers of the discriminator use filter kernels of size 512 with stride 1.

**Datasets**
We input the registration optical super-resolution images and SEM images to the neural network as LR/HR image pairs and crop them to 512×512 pixels. In order to make training faster and more efficient, the input image size is resized to 256*256 pixels. The training set has 600 images pairs. All the network output images shown in this paper were blindly generated by the deep network; that is, the input images had not previously been seen by the network. However, if such training image pairs are not available when using our super-resolution image transformation framework, an existing trained model could be used, although this might not produce ideal results in all cases.

**Strategy to avoid overfitting**
To prevent a situation in which the model performs well on the training set but only achieves mediocre performance on the cross-validation test, a data augmentation strategy is adopted to increase the model's capacity to generalize to unknown samples and avoid overfitting. For data augmentation, the original images are flipped horizontally and vertically and then rotated by 30°, 45°, 60°, and 90°. In addition, to increase the size of the dataset, we scale the original images to various sizes. To avoid the overfitting problem caused by the model having too many parameters and being too complex, the network design could include dropout in the training process to weaken the joint adaptation between neural nodes, which can prevent the synergy of certain features and make the model not too dependent on certain local features, thereby enhancing its robustness.

**Strategies to prevent artifact generation**
The data augmentation approaches discussed above also play a crucial role in overcoming artifacts by improving the mapping relationships between image pairings, thereby overcoming artifact generation. However, the presence of noise in the optical super-resolution images might limit the method's success in the image-to-image mapping of the super-resolution algorithm. To obtain sharper and clearer SEM images, noise reduction and image edge and contrast enhancement are performed during the pre-processing phase of the image dataset preparation. The design of the GAN model based on the residual network also plays a role in denoising. The deep residual network contains multiple residual blocks, skip connections, and block regularization, all of which provide excellent training outcomes. This enables the generation network to generate a noise-free image and the discriminator network to distinguish the generated image from the real image using the noise-free image as a reference, thus improving the quality of the generated image and effectively suppressing the occurrence of image artifacts. Because the image pairings in this study are manually aligned, some image matching errors may occur; hence, the network's loss function design does not include such overall image evaluation loss functions as PSNR or SSIM. Instead, the loss function used in ref [34] is applied. If the image pairings in the dataset can be properly aligned to create accurate mapping between images, adding PSNR or SSIM loss functions to generate superior super-resolution generation outcomes may be explored.

**Conclusion**
The diffraction limit in optical microscopy makes it extremely challenging to image and manipulate objects with a length scale below ~200 nm. SEM is often used for micro- and nano-scale structure observation, but the system is expensive, and the imaged sample must be oil-free and electrically conductive. Hence, to overcome the imaging procedural complexity, cost, and limitations for selected samples of SEM's, this work combines microsphere-assisted microscopy and deep learning methods to obtain high-contract and large depth-of-field SEM-like images for observing sub-diffraction limit features. Measured by PSNR and SSIM values, the deep learning method performs well in image-to-image translation, with a peak signal-to-noise ratio improvement of about 0.74 dB over the optical super-resolution image, which makes the visual features of the reconstructed image more detailed and natural. The experimental results show that the super-resolution imaging method based on deep learning and the SSUM system proposed in this paper has strong overfitting resistance, high training efficiency, and good detail performance in image reconstruction, indicating that it has broad applicability to chip detection, medical image analysis, remote sensing imagery, and a variety of other fields.

Using a combination of super-lens scanning microscopy imaging and deep learning, this work tackles several important issues in current image super-resolution-based reconstruction techniques. For example, the traditional deep learning-based single-input super-resolution image reconstruction methods are unable to achieve super-resolution image transformation from SSUM images to SEM images. Therefore, it is difficult to overcome the limitations of the production process and cost of imaging devices with nanoscale features. In contrast, using the technique described in this paper, super-resolution images of the same quality and resolution as SEM images can be generated directly from microscopic images. Moreover, the super-resolution images generated by the super-resolution reconstruction method using GAN is too smooth and loses the high-frequency information of the images, resulting in a texture structure lacking in richness and accuracy. In this work, images acquired by super-resolution techniques can learn and generate the same texture features as observed with SEM, resulting in images with better texture detail and more precise brightness and contrast information.

Therefore, our deep learning technique enables the generation of super-resolution images directly from images acquired by optical super-resolution devices while using fewer input images. We also demonstrate quantitatively that deep learning can achieve resolutions comparable to traditional SEM observed images. In summary, our work is crucial for the generalization of new high-resolution imaging systems and modalities and an essential step forward in the area of optical super-resolution and deep learning-based super-resolution imaging. This approach has the potential to transcend laboratory restrictions and allow for novel biological observations. We use a deep learning architecture that excels at image-to-image translation to close the gap between optical low-resolution and SEM high-resolution imaging systems.

### Acknowledgements

This work was supported by the Hong Kong Research Grants Council (Project No. 11216120) and the Science, Technology and Innovation Commission of Shenzhen Municipality (Grant Nos. SGDX2019081623121725 and JCYJ20190808181803703).

### Conflict of Interest

The authors declare no conflict of interest.

### Author contributions

H. S., H. L., F. W. contributed equally to this work. W. J. L., G. Z., and L. L. supervised and guided the project. W. J. L. and G. Z. conceived the initial concept of this work. M. C. and X. W. carried out the initial background review. W. J. L., H. S., H. L., H. Y. and G. Z. conceived the experimental design. F. W., W. J. L., and L. L. conceived the design of the SSUM system. H. L., X. W., and H. Y. used the SSUM system to acquire the SSUM images used in this work. H. L., Q. C., M. C., and X. W. helped with data collection. H. S. and G. Z. performed the experiments and analyzed the data with help from M. C., Q. C., X. W., and F. W., H. S and G. Z developed the deep learning algorithms with input from J. W. and D. W. H. S., H. L., G. Z. and W.J.L. co-drafted the manuscript, with input from F. W., J. W., and D. W., H. Y., J. W., D. W., L. L., G. Z., and W. J. L. provided critical technical advice and comments for the experimental studies and manuscript revision.